\documentclass[prb,floatfix,twocolumn,showpacs,superscriptaddress]{revtex4}
\usepackage{graphicx}% Include figure files
\usepackage{dcolumn}% Align table columns on decimal point
\usepackage{bm}% bold math
\usepackage{epsfig,psfrag,amsmath,amssymb,float}
\input{epsf}

\newcommand{\bc}{\begin{center}}
\newcommand{\ec}{\end{center}}
\newcommand{\be}{\begin{equation}}
\newcommand{\ee}{\end{equation}}
\newcommand{\beqn}{\begin{eqnarray}}
\newcommand{\eeqn}{\end{eqnarray}}

\begin{document}

\title{Crossover effects in the random exchange spin-$\frac{1}{2}$ antiferromagnetic chain}

\author{Nicolas Laflorencie}

\affiliation{Laboratoire de Physique Th\'eorique, CNRS-UMR5152
Universit\'e Paul Sabatier, F-31062 Toulouse, France}
\author{Heiko Rieger}
\affiliation{Theoretische Physik; Universit\"at des Saarlandes; 66041 Saarbr\"ucken; Germany}

\author{Anders W.~Sandvik}
\affiliation{Department of Physics, {\AA}bo Akademi University,
Porthansgatan 3, FIN-20500 Turku, Finland}

\author{Patrik Henelius}
\affiliation{Condensed Matter Theory, Physics Department, KTH, SE-106 91 Stockholm, Sweden}

\date{\today}

\begin{abstract}
The random antiferromagnetic spin-1/2 XX and XXZ chain is studied 
numerically for varying strength of the disorder, using exact
diagonalization and stochastic series expansion methods. The spin-spin
correlation function as well as the stiffness display a clear
crossover from the pure behavior (no disorder) to the infinite
randomness fixed point or random singlet behavior predicted by the the
real space renormalization group. The crossover length scale is shown to 
diverge as $\xi\sim{\mathcal D}^{-\gamma}$, where ${\mathcal D}$ is the 
variance of the random bonds. Our estimates for the exponent $\gamma$ agrees
well within the error bars with the one for the localization length
exponent emerging within an analytical bosonization calculation. Exact
diagonalization and stochastic series expansion results for the string
correlation function are also presented.
\end{abstract}

\maketitle

%%%%%%%%%%%%%%%%%%%%%%%%%%%%%%%%%%%%%%%%%%%%%%%%%%%%%%%%%%%%%%%%%%%%%%%%%%%%%%%
%%%%%%%%%%%%%%%%%%%%%%%%%%%%%%%%%%%%%%%%%%%%%%%%%%%%%%%%%%%%%%%%%%%%%%%%%%%%%%%

\section{Introduction}

Quantum spin chains exhibit a number of interesting features,
especially at low temperature when quantum fluctuations are stronger
than thermal ones. The antiferromagnetic (AF) Heisenberg model in one
dimension (1D) has been extensively studied since the discovery in
1931 of the Bethe Ansatz \cite{Bethe31} for the spin $S=\frac{1}{2}$
chain. In 1D, the AF XXZ model defined by the Hamiltonian
\be
\label{DefHamXXZ}
{\mathcal{H}}^{XXZ}=J\sum_{i=1}^{L}\Bigl[S_i^xS_{i+1}^x+S_i^yS_{i+1}^y+\Delta
S_i^zS_{i+1}^z\Bigr]
\ee
with $J>0$, exhibits a gap-less excitation spectrum for $\Delta \in
[-1,1]$ for $S=\frac{1}{2}$ (and more generally for half integer
spins~\cite{spins3.2}), whereas a gap opens up in the spectrum for
integer spins~\cite{Haldane83}.  In 1D, the quantum fluctuations
prevent the formation of long-range order~\cite{Mermin66} but
the correlation length of the model [Eq.(\ref{DefHamXXZ})] is
infinite and a {\it {quasi}}-long-range order (QLRO) emerges, with
power-law decaying spin-spin correlation functions in the ground state
(GS) :
\be
\label{DefCor}
C^{\alpha}(r)= \langle S^{\alpha}_{i}S^{\alpha}_{i+r}\rangle_{\rm{_{GS}}}
\propto\frac{(-1)^{r}}{r^{\eta_{\alpha}}}~{\rm{for}}~r\to \infty,
\ee
where $\alpha=x,y$ or $z$, $\langle...\rangle_{\rm{_{GS}}}$ is the GS
expectation value, and the critical exponent
$\eta_{x,y}=\eta_{z}^{-1}=1-\mu/\pi$, with $\mu=\arccos\Delta$.

If the AF exchange couplings are position-dependent, or more generally
distributed randomly according to a probability distribution
${\mathcal{P}}(J)$, the situation changes dramatically. Indeed, the
spin-$\frac{1}{2}$ chain described by the random-exchange XXZ
Hamiltonian
\be
\label{DefReXXZ}
{\mathcal{H}}_{\rm{random}}^{XXZ}=\sum_{i=1}^{L}
\Bigl[J_{\perp}(i)(S_i^xS_{i+1}^x+S_i^yS_{i+1}^y)
+J_{z}(i)\Delta S_i^zS_{i+1}^z\Bigr],
\ee
has lost the translation symmetry and rare events in the chain
dominate the low energy physics.\cite{Fisher94,Igloi00} Note that the
energy scale is set to unity by choosing mean values of random
couplings equal to 1. For instance, (i) the {\it random planar
exchange} (RPE) model with $J_{z}(i)=1,
\forall i$ and $J_{\perp}(i)$ random; (ii) the {\it{random z-z
exchange}} (RZE) model with $J_{\perp}(i)=1, \forall i$ and $J_{z}(i)$
random; (iii) the {\it{random exchange}} (RE) XXZ antiferromagnet for
which $J_{\perp}(i)=J_{z}(i)$ and are all random numbers.  

For the AF XXZ spin-$\frac{1}{2}$ chain, it has been shown by Doty and
Fisher~\cite{Doty92} that disorder is relevant and that any amount of
randomness destroys the QLRO and drives the system from a line of pure
fixed points to an infinite randomness fixed point (IRFP)
\cite{Fisher94}. The situation for higher spins $S>\frac{1}{2}$ is
more complicated since it depends \cite{HigherSpins} on the parity of
$2S$ and some issues are still under debate.\cite{Debate,SSESpins1}.  
Regarding the thermodynamic
properties of the random spin-$\frac{1}{2}$ XXX antiferromagnet, a
real space renormalization group (RSRG) scheme, introduced first by
Ma, Dasgupta and Hu~\cite{MDH} lead to a number of analytical results.
In particular, independent of the initial distribution
${\mathcal{P}}(J)$ of couplings the low energy properties at
the IRFP are characterized by a dynamical exponent $z=\infty$ and a GS
which consists of a tensorial product of randomly long-range coupled
dimers, the so-called {\it random-singlet phase}
(RSP)~\cite{Fisher94}.  In such a phase, the disorder averaged
spin-spin correlation function is dominated by strongly correlated
pairs and is therefore slowly decreasing, as a power-law
\be
\label{DefCorrRSP}
C^{\alpha}_{avg}(r)= {\overline{<S^{\alpha}_{i}S^{\alpha}_{i+r}>}}_{\rm{_{GS}}}\propto \frac{(-1)^{r}}{r^{\eta_{RSP}}},
\ee
where $\eta_{RSP}=2$ for all spin components $\alpha$ ($\Bar{\cdots}$
denotes the average over the disorder and the sites $i$). On the other
hand, in the RSP the {\it{typical}} correlations decay faster (i.e.\
with a stretched exponential) than the {\it{average}} correlations.
%
%\be
%\label{DefTypRSP}
%C^{\alpha}_{typ}(r)=e^{{\overline{\ln|<S^{\alpha}_{i}S^{\alpha}_{i+r}>_{\rm{_{GS}}}|}}} \propto e^{-A\sqrt{r}},
%\ee
%
%where $A$ is some nonuniversal constant.
These analytical predictions, that we will recall in greater detail in
Section 2, have been tested numerically several times using different
methods.  First, Lanczos exact diagonalization (ED) computations for
the model (\ref{DefReXXZ}) with RPE by S. Haas et al \cite{Haasetal}
for small chains ($L_{max}=18$) with the Hamiltonian (\ref{DefReXXZ})
with RPE appeared to confirm the universal behavior predicted in
Eq.(\ref{DefCorrRSP}), at least for strong disorder. Using
free-fermion ED, the RE XX model with the couplings uniformly
distributed over the interval $[0,1]$, has been investigated in
Refs.~\onlinecite{Henelius98,Igloi00} for large systems (up to $1024$ spins)
and the critical behavior of the correlation functions was found to
agree well with the analytical predictions.\cite{noteXX} Moreover, a
numerical RSRG calculation using again a uniform coupling distribution
has also confirmed theses predictions for the RE XXX
model \cite{Sigrist99} and other features of the RSP have been checked
at finite temperature by quantum monte carlo (QMC) simulations on
finite chains ($L_{max}=96$) in Ref~\onlinecite{Todo99}.  

Whereas the universal behavior described by Eqs.(\ref{DefCorrRSP})
appears o be well established, the recent density matrix
renormalization group (DMRG) calculations \cite{Stolze2002} for chains
(with free boundary conditions) defined by Eq.(\ref{DefReXXZ}) with
RPE caused a debate, \cite{ourcomment03} which we intend to settle in
this paper. Indeed, the conclusions of the DMRG simulations presented
in Ref.~\onlinecite{Stolze2002} on systems up to $400$ spins, quite similar 
to a previous one using smaller systems,\cite{Stolze96} disagree with the
IRFP scenario in so far as a dependence of the exponent $\eta$ upon
$\Delta$ and the disorder strength was claimed. The RSRG scheme is
expected to be asymptotically exact, but finite size (FS) effects
cannot be negligible, especially for weak disorder, i.e., far away from
the IRFP.  Indeed, one can show that the RG flow toward the IRFP is
controlled by a crossover characterized by a length scale $\xi$ which
is disorder dependent and diverges when the disorder strength is
approaching zero. Such a behavior has already been mentioned in,
\cite{ourcomment03} supported by exact diagonalization calculations
for large system sizes (up to $4096$ spins).

In this paper, we intend to shed more light on this disorder induced
crossover in random quantum spin-$\frac{1}{2}$ chains and
demonstrate convincingly via numerical studies of several related
models defined by Eq.(\ref{DefReXXZ}) the consistency of FS effects
and the IRFP scenario. Such a crossover from pure to random critical
behavior is very common in disordered systems and is always relevant,
in experiments as well as in numerical studies, when the disorder is not
too strong and the length scales that can be explored are not too
large. A good understanding of the order of magnitude of the crossover
length scale, in particular its scaling behavior in dependence of the
disorder strength, is therefore necessary in order not to be misled by
the mere appearance of the experimental and/or numerical data
(c.f.~Refs.\onlinecite{Stolze2002,ourcomment03}).

The paper is organized as follows. In Sec.~II, we first recall the
analytical predictions of the RSRG scheme. Then, using the
bosonization study of the weakly disordered spin-$\frac{1}{2}$ chain,
\cite{Doty92} we establish a disorder-dependent length scale which is
the localization length of the related problem of disordered particle
in 1D.\cite{Giamarchi88} We will see that this length scale is
closely related to the crossover length scale that emerges in our
calculations. In Sec.~III, we present the free-fermions ED results
for the spin-spin correlation functions for various disorder strengths
for system sizes (with periodic boundary conditions) up to $4096$
sites. The crossover length scale, which emerges naturally from the
data analysis, is studied as a function of the strength of the
disorder and compared with the localization length. The Ising part of
the Hamiltonian [Eq.(\ref{DefReXXZ})] has also been included and
investigated via QMC simulations, using the stochastic series
expansion (SSE) method. Section 4 is devoted to SSE calculations
performed at very low temperatures for the RE XXX model and the RPE
model.  First a brief explanation of the method is given and some
technical issues about equilibration and GS convergence are discussed;
then results for spin-spin and string correlation functions are
shown. Finally in section 5, we give a summary and some concluding
remarks.

\section{Analytical predictions}
\subsection{Real space renormalization group results}

\begin{figure}
\bc 
\epsfig{file=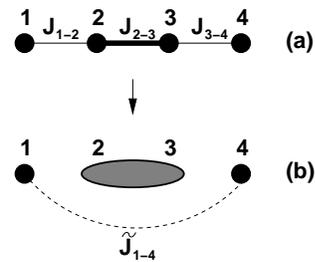,width=4cm,clip} 
\caption{Schematic picture of the decimation procedure for a 4-spins 
cluster. (a) The strongest coupling, between spins 2 and 3, is
decimated $\to$ (b) The spins 2 and 3 are frozen into a singlet and
the spins 1 and 4 are coupled by an effective coupling
${\tilde{J}}_{1-4}$ (see the text).}
\label{Pict4spins} 
\ec
\end{figure}

The RSRG method, introduced originally by Ma, Dasgupta and Hu for the
RE XXX spin-$\frac{1}{2}$ chain~\cite{MDH} has been developed and
studied in an exhaustive way by D. S. Fisher~\cite{Fisher94} for more
general RE XXZ Hamiltonians.  The basic ingredient of this decimation
procedure is a successive decrease of the energy scale via a successive
decimation of the strongest couplings in the chain.  Starting with a
$4$-spin cluster (see Fig.\ref{Pict4spins}) in which the exchange
$J_{2-3}$ is much larger than its neighbors $J_{1-2}$ and $J_{3-4}$,
the spins $2$ and $3$ are frozen into a singlet and the spins $1$ and
$4$ are coupled via an effective coupling
\be
{\tilde{J}}_{1-4}=\frac{J_{1-2}J_{3-4}}{2J_{2-3}}, 
\ee
calculated in second order perturbation expansion.

%%%%%%%%%%%%%%%FIG:PICT4SPINS%%%%%%%%%%%%%%%%%%%%%%%%%%%
%%%%%%%%%%%%%%%%%%%%%%%%%%%%%%%%%%%%%%%%%%%%%%%%%%
In the limit of infinite system size, Fisher has demonstrated the
existence of a fixed point for the the distribution of the effective
couplings, independent of the initial distribution, which is given by
\be
\label{IRFPDist}
{\mathcal{P}}_{0}({\tilde{J}})\propto {\tilde{J}}^{-1+\delta^{-1}},~\delta\to\infty.
\ee
Such a distribution is very broad and guarantees that the RSRG procedure
is asymptotically exact since the decimation produces weaker and
weaker couplings. The IRFP, characterized by the distribution
(\ref{IRFPDist}), is attractive for any amount of randomness in the
case of spins $\frac{1}{2}$ and the RSP, discussed in Section 1,
describes the GS. At the critical point, the energy and length scales
are related via
\be
\label{ScalGap}
\ln \Delta_{\epsilon} \sim -\sqrt{L}
\ee
and as a consequence, the dynamical exponent $z$ is
infinite. Concerning the correlation functions, as already mentioned
in Section 1, the average and typical values behave quite differently
since rare events control the physics (see
Eq.~[\ref{DefCorrRSP})].  The average correlation
function is dominated by long-range paired singlet and takes the
following expression, independently of the direction (transverse or
longitudinal)
\be
C_{avg}(r)\propto \frac{(-1)^{r}}{r^{2}}.
\ee

Another quantity, which measure an hidden order is the 
{\it{string correlation function}}, defined in the GS at distance $r$ by 
\be
S(r)=\frac{2^{r+1}}{L}\sum_{i=1}^{L}<S_{i}^{z}S_{i+1}^{z}...S_{i+r}^{z}>_{\rm{_{GS}}}.
\label{string}
\ee
At the IRFP, the disorder averaged expectation value $S_{avg}(r)$ is
expected to decrease as a power law, with a well-defined
exponent:\cite{Fisher92.95}
\be
S_{avg}(r)={\overline{S(r)}}\propto \frac{(-1)^{r}}{r^{2-\phi}},
\label{string_exp}
\ee
$\phi=\frac{1+\sqrt{5}}{2}$ being the golden mean. The RSRG method
yields several other properties at the IRFP ; in particular finite
temperature properties are well understood, but it is not the aim of
our work.

To give a brief conclusion for this part, we could say
that this real space decimation procedure is analytically solvable for
the random spin-$\frac{1}{2}$ chain and is asymptotically
exact. Nevertheless, this theory does not quantify the FS effects,
which are always present in numerical simulations. Therefore, in order
to give good interpretations of numerical results, the understanding
of FS effects is crucial. This is what we are striving for here, 
using the bosonization treatment of the random chain.

%%%%%%%%%%%%%%%%%%%%%%%%%%%%%%%%%%%%%%%%%%%%%%%%%%%%%%%%%%%%%%%%%%%%%%%%%%%%%%%
\subsection{Bosonization of the random chain: 
Emergence of a disorder-dependent length scale}

In this part, we summarize previous results obtained using
bosonization techniques.\cite{Giamarchi88,Doty92} The XXZ
spin-$\frac{1}{2}$ chain can be mapped using the Jordan-Wigner
transformation (see Section 3) into a spinless interacting fermions
problem in 1D.  The low-energy excitations around the Fermi points can
be considered in terms of bosonic fields and the resulting Hamiltonian
describes a Luttinger liquid.\cite{review00} It is characterized by a
set of $\Delta$-dependent {\it{Luttinger liquid parameters}} which are
the velocity of excitations $u$ and the parameter $K$, given by
\begin{eqnarray}
u(\Delta)=\frac{\pi}{2} \frac{\sin(\mu)}{\mu},\nonumber\\
K(\Delta)=\frac{\pi}{2(\pi-\mu)}.
\end{eqnarray}
Several types of quenched randomness added to the pure XXZ model has
been studied by Doty and Fisher in \cite{Doty92} where they found, for
random perturbations that preserves the XY symmetry, a critical
behavior which belongs to the universality class of the
Giamarchi-Schulz transition for 1D bosons in a random
potential.\cite{Giamarchi88}  Let us define the disorder strength
${\mathcal{D}}$ by
\be
\label{DisStr}
{\mathcal{D}}_{\perp ,z}={\overline{(J_{\perp ,z}(i))^2}}-\Bigl({\overline{J_{\perp ,z}(i)}}\Bigr)^2.
\ee
More precisely, for the RPE model ${\mathcal{D_{\rm
RPE}}}={\mathcal{D_{\perp}}}$, for the RZE model, ${\mathcal{D_{\rm
RZE}}}={\mathcal{D}}_{z}$ and for the RE XXZ model, since the
randomness is isotropic ${\mathcal{D_{\rm
RE-XXZ}}}={\mathcal{D}}_{\perp}={\mathcal{D}}_{z}$.  For a weak random
perturbation added to the planar exchange, the renormalization under a
change of length scale $l=\ln L$ is
\be
\label{RG}
\frac{{\partial}{\mathcal{D}}}{{\partial}l} = (3-2K){\mathcal{D}}.
\ee
Therefore, if $K<3/2$ (i.e. $-\frac{1}{2}<\Delta<1$) the disorder is a
relevant perturbation and the phase is the RSP. The renormalization
flow toward the IRFP is controlled by a length scale which emerges
from Eq.(\ref{RG}) :
\be
\label{xi}
\xi^*({\mathcal{D}})\sim {\mathcal{D}}^{-\frac{1}{3-2K}}.
\ee
For the RE XXX model, the random perturbation added to the operator
$S_{i}^{z}S_{i+1}^{z}$ is marginally irrelevant and therefore the
exponent ${\frac{1}{3-2K}}=\frac{1}{2}$ for $\Delta=1$ is expected to
have small logarithmic corrections~\cite{Orignac}.

The length scale $\xi^*$ is called the localization length since in
the fermionic language, the transition at ${\mathcal{D}}>0$ is a
localization transition.\cite{Giamarchi88,anderson}  Such a
metal-insulator transition driven by the disorder is characterized for
instance, by the vanishing of the zero temperature Drude weight (also
called the charge stiffness or the spin stiffness in the case of the
spin-$1/2$ XXZ model) $\forall {\mathcal{D}}>0$ in the thermodynamic
limit.  Previous numerical studies have checked this effect on
transport properties in the case of interacting spinless fermions in a
random potential using ED~\cite{Poilblanc94} or equivalently in the
case of the XXZ chain in a random magnetic
field.\cite{Runge94,Urba03}  For the simpler model of non-interacting
fermions with random hopping, mapped into the RE XX spin chain, very
large scale numerical simulations have been carried out on systems up
to $2048$ sites.\cite{Stif03} Using the scaling law for the spin
stiffness

\be
\label{ScalStif}
\rho_{s}(L,{\mathcal{D}})=g(L/\xi^*({\mathcal{D}})),
\ee

the localization length has been precisely studied and agrees
perfectly with Eq.(\ref{xi}) for weak disorder (see Section 3 C and
Figs.\ref{fig:STI.SCALE}-\ref{fig:XISTI}).  Our purpose here is to
study crossover effects for various 1D spin-$\frac{1}{2}$ disordered
models. Regarding the low-energy effective theory predicted by some
bosonization calculations, there is only one relevant length scale
which emerges from it, i.e., the localization length
$\xi^*({\mathcal{D}})$.  Based on numerical calculations performed
over FS clusters for various disorder strengths, the next sections are
dedicated to the study of the disorder dependence of the crossover
length scale and its comparison with the localization length.
 
%%%%%%%%%%%%%%%%%%%%%%%%%%%%%%%%%%%%%%%%%%%%%%%%%%%%%%%%%%%%%%%%%%%%%%%%%%%%%%%%%%%%%%%%%%%%%%%%%%%%%%%%%%%%%%%%%%%%%%%%%%%%%%%%%%%%%%%%%%%%%%%%%%%%%%%%%%%%%%
\section{Exact diagonalization study at the XX point}

\subsection{Free fermions representation}

Let us consider the 1D XX spin $\frac{1}{2}$ model with random exchange couplings $J_{\perp}(i)$. 
This quantum problem is governed by the following lattice Hamiltonian
\begin{equation}
\label{DefReXX}
{\mathcal{H}}_{\rm{random}}^{XX}=\sum_{i=1}^{L}\Bigl[J_{\perp}(i)(S_i^xS_{i+1}^x+S_i^yS_{i+1}^y)\Bigr].
\end{equation}
We impose periodic boundary conditions;
$\vec{S}_{L+1}=\vec{S}_{1}$.  It's well known that this spin problem
can be mapped into a free spinless fermions model via the
Jordan-Wigner transformation~:~$S^z_j = 1/2-n_j$, and $S^+_j
=c_je^{i\pi\sum_{l=1}^{j-1}n_l}$.  The $c_j$ satisfy fermionic
commutation relations, $ \lbrace c^\dagger_i,c_j\rbrace=\delta_{ij}$,
and $n_j = c^\dagger_jc_j$ is the number of fermions (spin down) at
the j-site. The Hamiltonian can then be written as
\begin{eqnarray}
\label{DefReXXFerm}
{\mathcal{H}}_{\rm{random}}^{XX} = \sum_{i=1}^{L-1}\Bigl[\frac{J_{\perp}(i)}{2}(c_i c_{i+1}^{\dagger}  +  h.c)\Bigr]~\nonumber \\
+\frac{J_{\perp}(L)}{2}e^{i\pi \mathcal{N}}(c_L c_{1}^{\dagger}  +  h.c), 
\end{eqnarray}
where $h.c$ is the hermitian conjugate and $\mathcal{N}=\sum_{i=1}^{L}
n_{i}$ is the number of fermions in the system. In the non random
case, the solution of the problem via a Fourier transformation is
trivial~\cite{Lieb61} because of the translational invariance.  But
in the random system, this symmetry is broken and we have to solve
numerically a random matrix problem.  The way to obtain the correlation
functions is straightforward and has already been explained in several
previous works;\cite{Lieb61,Young96,Henelius98,Igloi00} it amounts
to a numerical calculation of the eigenvectors of a $L
\times L$ band matrix and then the evaluation of a $(r-1)\times (r-1)$
(resp. $2\times 2$) determinant in order to compute the transverse
(resp. longitudinal) spin-spin correlation function at distance $r$
$C^x(r)$ (resp. $C^z(r)$). We can note that in the same way, the
string correlation functions can also be obtained.\cite{Henelius98}

%%%%%%%%%%%%%%%%%%%%%%%%%%%%%%%%%%%%%%%%%%%%%%%%%%%%%%%%%%%%%%%%%%%%%%%%%%%%%%%
\subsection{Numerical results for the spin-spin correlation functions: crossover effects}

%%%%%%%%%%%%%%%FIG:CORRXX%%%%%%%%%%%%%%%%%%%%%%%%%%%
\begin{figure}
\bc 
\epsfig{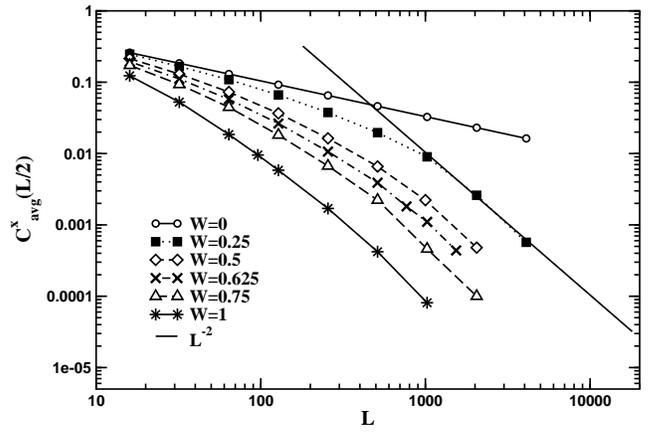} 
\caption{Averaged correlation function $C^{x}_{avg}(L/2)$ as a function of the system size $L$ on a log-log
scale for $W=0, 0.25, 0.5, 0.625, 0.75, 1.0$ (from top to bottom). The data are averaged over 50000 samples for $L \le 1024$, 3000 for $L=2048$ and 500 for $L=4096$, the statistical errors are smaller than the symbol sizes. The data for the pure system ($W=0$) follow $C^{x}(L/2)\propto L^{-1/2}$, the full line with slope $-2$ is the expected asymptotic behavior according to the IRFP scenario.}
\label{fig:CORRXX} 
\ec
\end{figure}
%%%%%%%%%%%%%%%%%%%%%%%%%%%%%%%%%%%%%%%%%%%%%%%%%%

\begin{figure}
\bc 
\epsfig{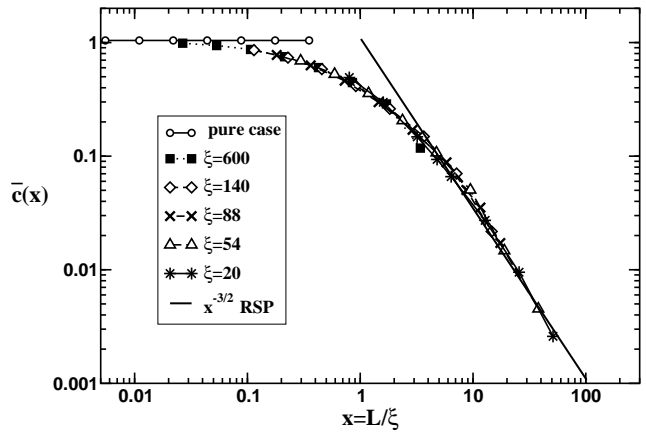} 
\caption{Scaling plot according to
  Eq. (\ref{ScalXX}) of the data shown in Fig.\ref{fig:CORRXX} with
  $\xi=600$, $140$, $88$, $54$, $20$ for $W=0.25$, $0.5$, $0.625$,
  $0.75$ and $1.0$, respectively. The symbols are identical to
  Fig.\ref{fig:CORRXX}.}
\label{fig:ScalXX} 
\ec
\end{figure}

In order to study the crossover as a function of the disorder
strength, we have chosen the following $W$-dependent flat bond
distribution
\be
\label{DistDis}
{\mathcal{P}}(J_{\perp})=\left\{
\begin{array}{rl}
\frac{1}{2W} & \mathrm{if}\ J_{\perp}\in [1-W,1+W]\\
0 & {\mathrm {otherwise}}\
\end{array}
\right.
\ee
The disorder strength, defined by Eq.(\ref{DisStr}), is
${\mathcal{D}}=W^2/3$ and we define $\delta$ as the variance of the
random variable $\ln J_{\perp}(i)$ by
\be
\label{delta}
\delta^2={\overline{(\ln J_{\perp}(i))^{2}}}-\Bigl({\overline{\ln J_{\perp}(i)}}\Bigr)^{2}
\ee
which is related to $W$ according to
\be
\delta={\sqrt{1-\frac{1-W^2}{4W^2}\Bigl[\ln\Bigl(\frac{1+W}{1-W}\Bigr)\Bigr]^2}}.
\label{relwdel}
\ee
We note that
for weak disorder $W \ll 1$, $\delta \sim \sqrt{D}$.  In order to
reduce statistical errors and boundary effects we have used the PBC
and computed the bulk correlation function in the transverse direction
at mid-chain
\be
C^{x}(\frac{L}{2})=\frac{2}{L}\sum_{i=1}^{\frac{L}{2}}
\langle S_{i}^{x}S_{i+\frac{L}{2}}^x\rangle_{\rm{_{GS}}}
\ee
for several system sizes ($L=2^q,~q=1,...,12$) and disorder strengths
($W=0.25,~0.5,~0.625,~0.75,~1$). The data for $C^{x}(\frac{L}{2})$
were computed for each individual sample using standard routines and
then averaged over the disorder. The number of disorder configurations
was more than $5\cdot10^{4}$ for $L\le 1024$ and at least
$500$ for the largest size and weakest randomness ($L=4096$
$W=0.25$). In Fig.\ref{fig:CORRXX}, we show the average bulk
correlation function
$C^{x}_{avg}(\frac{L}{2})={\overline{C^{x}(\frac{L}{2})}}$ for
different disorder strengths.  We observe that for small system sizes
the slope of $C^{x}(\frac{L}{2})$ versus $L$ in a log-log plot is much
smaller than $2$, the value that one would expect form the IRFP
scenario. But when $L$ increases one observes a crossover, as reported in
Ref.~\onlinecite{ourcomment03}, from an apparently non-universal behavior 
with a $W$-dependent power law exponent $\eta(W)$ for small sizes to a
universal behavior with $C^{x}_{avg}(\frac{L}{2})\sim L^{-2}$ for
$L\to \infty$, as predicted by the RSRG.\cite{Fisher94}
 Such a behavior suggests the existence of a
disorder-dependent crossover length scale $\xi$ which controls the
crossover from the pure (instable) fixed point to the IRFP which is
attractive, even for weak disorder. Defining the dimensionless
parameter $x=L/\xi$, one can identify three different
regimes: \\ 
(i) For $x \ll 1$, the critical behavior of the pure
system ($J_{\perp}(i)={\rm constant}$) is dominant, with an exponent
$\eta(W)= 1/2$. \\ 
(ii) For $x \gg 1$, we are in the asymptotic
regime where the predictions of the RSRG are recovered, in particular
$\eta(W)=\eta_{_{RSP}}=2$. 
\\ (iii) For $x\sim 1$ we are in the
crossover regime with a $W$- and $L$-dependent effective (FS) exponent
$\eta(W)$. \\ 
Consequently, we expect the following scaling form:
\be
\label{ScalXX}
C^{x}_{avg}(\frac{L}{2})=L^{-1/2}{\tilde c}(L/\xi)
\ee
where the scaling function $\tilde c(x)$ is constant in the regime
(i), and $\tilde c(x)\to x^{-3/2}$ in the regime (ii). In
Fig.\ref{fig:ScalXX}, the scaling plot following Eq.(\ref{ScalXX}) is
shown for the data of Fig.\ref{fig:CORRXX}. $\xi(W=1)$ has been chosen
such that the crossover region (iii) is centered around $x\simeq 1$
and the other estimates have been adjusted in order to give the best
data collapse.

%%%%%%%%%%%%%%%FIG:SCALXX%%%%%%%%%%%%%%%%%%%%%%%%%%%
%%%%%%%%%%%%%%%%%%%%%%%%%%%%%%%%%%%%%%%%%%%%%%%%%%

%%%%%%%%%%%%%%%%%%%%%%%%%%%%%%%%%%%%%%%%%%%%%%%%%%%%%%%%%%%%%%%%%%%%%%%%%%%%%%%
\subsection{The crossover length scale as a localization length}

%%%%%%%%%%%%%%%FIG:XiXX%%%%%%%%%%%%%%%%%%%%%%%%%%%
\begin{figure}
\bc 
\epsfig{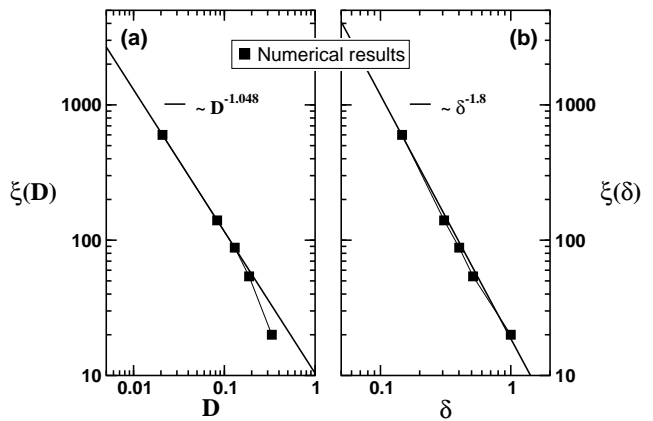} 
\caption{Disorder dependence of the crossover length scale $\xi$ of the 
random XX chain. The full squares are the numerical estimates from the
data collapse in Fig.\ref{fig:ScalXX} and full lines are fits. (a) In
function of the disorder parameter ${\mathcal{D}}$, a power law with
an exponent $-1.048$ fits the data only for weak disorder whereas in
(b), a fit $\xi(\delta)\sim \delta^{-1.8}$ works for the entire range
of disorder strength studied here.}
\label{fig:XiXX} 
\ec
\end{figure}
%%%%%%%%%%%%%%%%%%%%%%%%%%%%%%%%%%%%%%%%%%%%%%%%%%

In this part, the dependence of the crossover length scale on the
disorder strength is studied. A comparison with the localization
length $\xi^*$, calculated using the spin stiffness of the RE XX chain,
is also presented. Fig.\ref{fig:XiXX} shows a plot of $\xi$ vs
the disorder parameters ${\mathcal{D}}$ and $\delta$. As expected one
can observe a singular behavior for $\mathcal{D}~{\rm{or}}~\delta \to
0$.  More precisely, we observe in Fig.\ref{fig:XiXX}(a) that for
sufficiently weak disorder (typically for ${\mathcal{D}}<0.1$), the
crossover length scale is well fitted by a power-law :
$\xi({\mathcal{D}})\propto {\mathcal{D}}^{-\gamma}$ with an exponent
$\gamma =1\pm 0.1$, in good agreement with the localization length
exponent predicted by Eq.(\ref{xi}) which gives $\frac{1}{3-2K}=1$ at
the XX point.  For stronger disorder, a deviation from the power-law
is observed. On the other hand, $\xi(\delta)$ shown in
Fig.\ref{fig:XiXX}(b), can be fitted by a power-law
$\xi(\delta)\propto \delta^{-\Phi}$, with $\Phi=1.8 \pm 0.2$ for the
whole range of randomness studied here.

%%%%%%%%%%%%%%%FIG:STI.SCALE%%%%%%%%%%%%%%%%%%%%%%%%%%%
\begin{figure}
\bc 
\epsfig{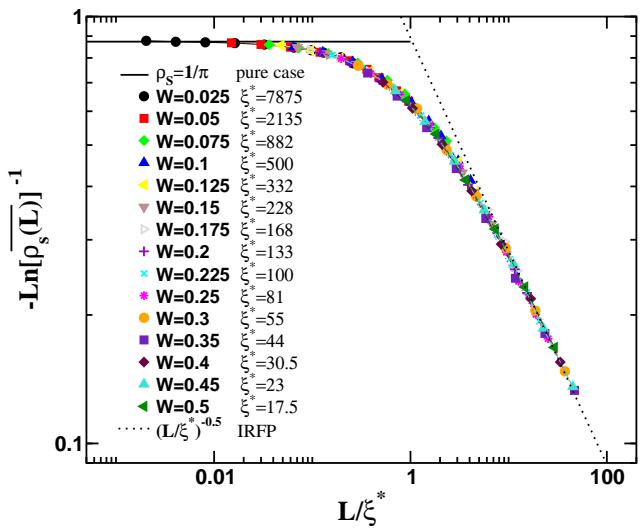} 
\caption{Inverse logarithm of the disorder averaged spin stiffness plotted for several box sizes $W$ specified on the plot. All the curves are collapsed since a rescaling of the $x$-axis has been done, providing an universal curve 
as a function of $L/\xi^*$. The $W$-dependent localization length $\xi^*$ has been calculated for each disorder strength, as indicated on the plot, in order to give the best data collapse. The full line stands for the pure case and the dotted one is for the IRFP behavior.}
\label{fig:STI.SCALE} 
\ec
\end{figure}
%%%%%%%%%%%%%%%%%%%%%%%%%%%%%%%%%%%%%%%%%%%%%%%%%%

It is instructive to compare the crossover length scale $\xi$ with the
localization length $\xi^*$, extracted from the numerical calculation
of the spin stiffness of the RE XX chain (for more details about this
calculation, see Ref.~\onlinecite{Stif03}).  While the transport properties of
random spin chains are not the purpose of this
paper,\cite{HuseTransport} we mention here some results that two
of us obtained by ED performed on the RE XX chain.\cite{Stif03}  The
spin stiffness $\rho_s$ which measures the magnetization transport
along the ring is calculated in the GS as the second derivative of the
GS energy per site with respect to a twist angle $\varphi$ applied at
the boundaries using the so-called {\it{twisted boundary
conditions}},\cite{TBC} and taking the limit $\varphi \to 0$.  For
the same model [Eq.(\ref{DefReXX})] studied in this section and for
systems sizes going from $8$ to $2048$ sites, $\rho_s$ has been
calculated by ED techniques for several disorder strengths (from
$W=0.025$ to $W=1$) and averaged over a very large number of samples
(from $10^5$ for the smallest sizes to $500$ for the largest).

%%%%%%%%%%%%%%%FIG:XISTI%%%%%%%%%%%%%%%%%%%%%%%%%%%
\begin{figure}
\bc 
\epsfig{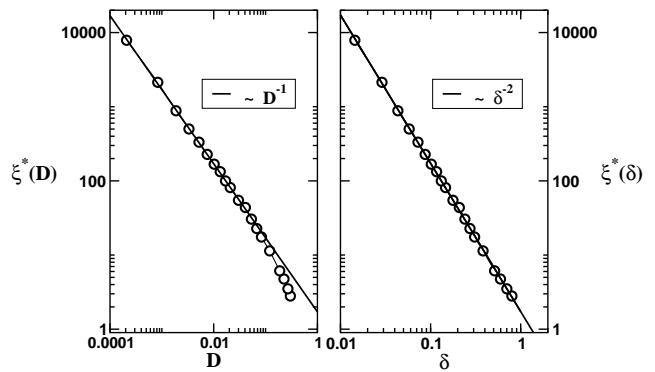} 
\caption{Disorder dependence of the localization length $\xi^*$ of the random XX chain calculated using the scaling of the stiffness [Eq.(\ref{ScalStif})]. (a) In function of the disorder parameter ${\mathcal{D}}$, the expected power-law Eq.(\ref{xi}) with an exponent equal to $-1$ is in perfect agreement with numerical data (open circles) which can be fitted for weak disorder, with an exponent equal to $-1\pm 0.01$. (b) In function of the other disorder parameter $\delta$, the numerical data (open circles) are perfectly described by a power-law, for the entire range of disorder, with an exponent equal to $-2\pm 0.02$.}   
\label{fig:XISTI} 
\ec
\end{figure}
%%%%%%%%%%%%%%%%%%%%%%%%%%%%%%%%%%%%%%%%%%%%%%%%%%

The stiffness $\rho$ has dimension of inverse (${\rm
length}^{d-2}\times\xi_\tau$), where $\xi_\tau$ is the correlation
length in the imaginary time direction.\cite{Wallin94} In our case
$\xi_\tau\sim\exp(A\xi^{1/2})$, which is one manifestation of the IRFP
that dominates the critical behavior of the random XX chain, and
$\xi=L$ for a finite system at criticality we expect $\rho$ to scale
as $\ln {\overline{\rho_S(L)}} \sim -{\sqrt{L}}$. Combining this
with Eq.(\ref{ScalStif}) we show in Fig.\ref{fig:STI.SCALE} a scaling
plot of $-(\ln g(L/\xi^*))^{-1}$ which displays the same features as
Fig.\ref{fig:ScalXX}.  Indeed, for $L\ll\xi^*$, the
pure behavior is observed with a stiffness $\rho_S
\simeq{1}/{\pi}$,\cite{noteSti} and for $L\gg\xi^*$, the IRFP
behavior is recovered with $\ln g(L/\xi^*) \sim
-(L/\xi^*)^{0.5}$; the regime where $L \sim \xi^*$ being a crossover
regime.

The localization length $\xi^*(W)$ has been estimated for different
values of the disorder strength (note that the computational demand
for calculating the stiffness is substantially smaller than the one
for the correlation function~\cite{noteNum}, for which reason we could
compute more data points) and is shown on the Fig.\ref{fig:XISTI}
versus the disorder parameter. We see clearly that the behavior of the
crossover length $\xi$ as a function of the disorder strength (see
Fig.\ref{fig:XiXX}) is exactly analogous to the one of the
localization length $\xi^*$.  Indeed, for $\mathcal{D} \ll 1$, the
bosonization result Eq.(\ref{xi}) agrees with numerical results, as
shown in Fig.\ref{fig:XISTI}(a), and for stronger disorder we observe
the same deviation as in $\xi({\mathcal{D}})$. 
Fig.~\ref{fig:XISTI}(b) gives us the confirmation
that for strong disorder the equation (\ref{xi}) has to be replaced by
\be
\label{xidelta}
\xi^*(\delta)\sim \delta^{-\Phi}.
\ee 

Since for weak disorder $\delta \sim \sqrt{\mathcal{D}}$, we expect:
$\Phi=\frac{2}{3-2 K}$ which works perfectly for the entire range of
disorder considered here, as shown in Fig.\ref{fig:XISTI}(b). 

Let us summarize our results that we obtained so far for the RE XX
chain.  With ED calculations we studied the crossover that controls
the renormalization flow starting from a system with a finite disorder
to an infinite disorder fixed point. As predicted by RSRG and
bosonization calculations, the IRFP is attractive for any amount of
initial disorder and the crossover length scale $\xi$ is well
described by a power-law, diverging like
${\mathcal D}^{-\gamma}$.  Moreover,
the exponent $\gamma$ has been identified to be identical with
the localization length exponent occurring in $\xi^*({\mathcal{D}})\sim
{\mathcal{D}}^{-\frac{1}{3-2 K}}$.  While the parameter ${\mathcal
{D}}$ is suitable to quantify the divergence near $0$, we have found the
parameter $\delta$, Eq.~(\ref{delta}), to be a better candidate to describe 
localization and/or crossover behaviors for any strength of randomness :
$\xi(\delta)\sim \xi^*(\delta) \propto \delta^{-\Phi}$ with
$\Phi=\frac{2}{3-2 K}$.

%%%%%%%%%%%%%%%%%%%%%%%%%%%%%%%%%%%%%%%%%%%%%%%%%%%%%%%%%%%%%%%%%%%%%%%%%%%%%%%
%%%%%%%%%%%%%%%%%%%%%%%%%%%%%%%%%%%%%%%%%%%%%%%%%%%%%%%%%%%%%%%%%%%%%%%%%%%%%%%
\section{Quantum Monte Carlo study}

\subsection{The SSE method}

The SSE QMC method has been described as a loop algorithm in detail by
one of us in.\cite{Sandvik.ref}  More recently the concept of
directed loop has been developed
\cite{SandvikDirect02,Syljuasen02,Alet03} and the efficiency of such
an algorithm has been demonstrated for several models, in particular
for the XXZ model, defined by Eq.(\ref{DefHamXXZ}).  We start from the
general random-exchange XXZ Hamiltonian (\ref{DefReXXZ}) that we can
rewrite as a sum over diagonal and off-diagonal operators
$$
{\mathcal{H}}_{\rm{random}}^{XXZ}=-\sum_{b=1}^{L}\Bigl[
J_{z}(b)H_{1,b}-J_{\perp}(b)H_{2,b}\Bigr],
$$ 
where $b$ denotes a bound
connecting a pair of interacting spins $(i(b),j(b))$.
\be
H_{1,b}=C-\Delta S^{z}_{i(b)}S^{z}_{j(b)}
\ee
is the diagonal part and the off-diagonal part is given by
\be
H_{1,b}=\frac{1}{2}[S^{+}_{i(b)}S^{-}_{j(b)}+S^{-}_{i(b)}S^{+}_{j(b)}],
\ee
in the basis
$\{|\alpha>\}=\{|S^{z}_{1},S^{z}_{2},...,S^{z}_{L}>\}$. The constant
$C$ which has been added to the diagonal part ensures that all
non-vanishing matrix elements are positive.  The SSE algorithm is
based on Taylor expanding the partition function
$Z=Tr\{e^{-\beta{\mathcal{H}}_{\rm{random}}^{XXZ}}\}$ up to a cutoff
${\mathcal{M}}$ which is adapted during the simulations in order to
ensure that all the elements of order higher than ${\mathcal{M}}$ in
the expansion do not contribute. So
\be
Z=\sum_{\alpha}\sum_{S_{\mathcal{M}}}\frac{\beta^{n}({\mathcal{M}}-n)!}{{\mathcal{M}}!}\Bigl<\alpha\Bigl|\prod_{i=1}^{\mathcal{M}}J_{a_i}(b_i)H_{a_i,b_i}
\Bigr|\alpha\Bigr>,
\ee
where $S_{\mathcal{M}}$ denotes a sequence of operator indices 
\be
\label{Op.Ind}
S_{\mathcal{M}}=[a_1,b_1],[a_2,b_2],...[a_{\mathcal{M}},b_{\mathcal{M}}]
\ee
with $a_i=1,2$ corresponds to the type of operator (diagonal or not) and $b_i=1,2,...L$ is the bond index. 
Note that $J_1(b)=J_z(b)$ and $J_2(b)=J_{\perp}(b)$. A Monte Carlo configuration is therefore defined by a state $|\alpha\rangle$ 
and a sequence $S_{\mathcal{M}}$.
Of course, a given operator string does not contain $\mathcal{M}$ operators of type $1$ or $2$, but only $n$; so in order to keep constant the size of $S_{\mathcal{M}}$, ${\mathcal{M}}-n$ unit operators $H_{0,0}=1$ have been inserted in the string, taking into account all the possible ways of insertions. 
The starting point of a simulation is given by a random initial state $|\alpha\rangle$ 
and an operator string containing ${\mathcal{M}}$ unit operators $[0,0]_1,...,[0,0]_{\mathcal{M}}$. 
The first step is the {\it{diagonal update}} which consists in exchanging unit and diagonal operators at each position $p$ $[0,0]_p\leftrightarrow [1,b_i]_p$ in $S_{\mathcal{M}}$ with Metropolis acceptance probabilities
\begin{eqnarray}
P_{[0,0]_p\rightarrow [1,b]_p}={\rm{min}}\left [1,\frac{J_{z}(b)L\beta\Bigl<\alpha(p)\Bigl|H_{1,b}\Bigr|\alpha(p)\Bigr>}{{\mathcal{M}}-n}\right ],\\
P_{[1,b]_p\rightarrow [0,0]_p}={\rm{min}}\left [1,\frac{{\mathcal{M}}-n+1}{J_{z}(b)L\beta\Bigl<\alpha(p)\Bigl|H_{1,b}\Bigr|\alpha(p)\Bigr>}\right ].
\end{eqnarray}
During the ``propagation'' from $p=1$ to $p={\mathcal {M}}$, the ``propagated'' state 
\be
|\alpha(p)\rangle\sim \prod_{i=1}^{p}H_{a_i,b_i}|\alpha\rangle
\ee
is used and the number of non-unit operators $n$ can varies at each index $p$. 
The following step is the {\it{off-diagonal update}}, also called the {\it{loop update}}, carried out at $n$ fixed. 
Its purpose is to substitute $[1,b_i]_p\leftrightarrow [2,b_i]_p$ in a cluster-type update, i.e., with the
operators forming closed loops.
Such a construction has already been discussed in detail elsewhere \cite{SandvikDirect02}. 
A very efficient directed loop implementation can be used and for $\Delta \in [0,1]$ it has been shown that during the construction of the loop, back-tracking processes can be avoided. 
At the $SU(2)$ AF point, the algorithm is deterministic because we can build all the loops in a unique way. 
So, for $\Delta=1$, all the loops are updated independently of each other with probability $1/2$. 
For $\Delta \neq 1$ the construction of the loop depends on some well defined probabilities \cite{SandvikDirect02} at each time a non unit operator is encountered in the loop building.\\
One MC step is consists of  {\it{diagonal}} updates at all possible locations in the index
sequence, followed by a number of loop updates (the number adjusted so that the average number of operators
changes is comparable to the total number of operators). 
Before starting the measurement of physical observables, one has to perform equilibration steps, notably necessary to adapt the cutoff ${\mathcal{M}}$. 

%%%%%%%%%%%%%%%%%%%%%%%%%%%%%%%%%%%%%%%%%%%%%%%%%%%%%%%%%%%%%%%%%%%%%%%%%%%%%%%

\subsection{Convergences issues}

%%%%%%%%%%%%%%%%%%%%%%%%%%%%%%%%%%%%%%%%%%%%%%%%%%%%%%%%%%%%%
\begin{figure}
\bc
\epsfig{file=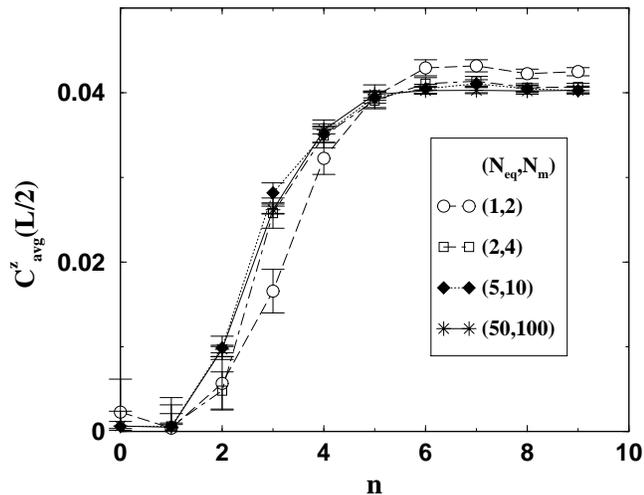,width=8.4cm,clip} 
\caption{Test for the convergence of the disorder averaged longitudinal correlation function calculated for a the RE XXX chain at $W=0.5$ with $16$ sites. Results, averaged over $10^3$ samples, are for different number of MC steps ($N_{eq},N_m$) as shown on the plot. The $\beta$-doubling scheme has been used with inverse temperatures $\beta_n=2^n$,
with $n$ used here for the $x$-axis.}
\label{Cz.Equil}
\ec
\end{figure}
%%%%%%%%%%%%%%%%%%%%%%%%%%%%%%%%%%%%%%%%%%%%%%%%%%%%%%%%%%%%%

The precise determination of physical observables using QMC suffers
obviously from statistical errors since the number of MC steps is
finite.  As we deal with disordered spin chains, the sample to sample
variation is another source of errors. Moreover, the calculation of GS
expectation values for a system close to an IRFP, where FS gap scale
like $\ln \Delta_{\epsilon} \sim \sqrt{L}$, requires a very carefully
numerical treatment.  In order to avoid finite temperature effects
and to ensure that we measure observable in the GS, we use
the $\beta$-doubling scheme, developed in Ref.~\onlinecite{Sandvik.perc} and
then used in Refs.~\onlinecite{SSESpins1,laflo03}.  Such a scheme is a very
powerful tool because it allows to reach extremely low temperatures
rather rapidly {\it{and}} reduces considerably equilibration times in
the MC simulation. The procedure is quite simple to implement and its
basic ingredient consists in carrying out simulations at successive inverse
temperatures $\beta_n=2^n$, $n=0,1,...,n_{max}$. Starting with a given
sample at $n=0$ we perform a small number of equilibration steps
$N_{eq}$ followed by $N_m=2 N_{eq}$ measurement steps.  At the end of
the measurement process, $\beta$ is doubled (i.e. $n \rightarrow n+1$)
and in order to start with an ``almost equilibrated'' MC
configuration, the starting sequence used is the previous
$S_{\mathcal{M}}$ doubled, i.e.,
\be
S_{2\mathcal{M}}=[a_1,b_1],...[a_{\mathcal{M}},b_{\mathcal{M}}][a_{\mathcal{M}},b_{\mathcal{M}}],...,[a_1,b_1].
\ee
Such a scheme becomes very efficient at low temperature and for disordered systems, in which very small correlations may develop. It is for the moment the most efficient technique available to cancel remaining temperature effects although a zero-temperature SSE algorithm might be developed soon.\cite{NoteTodo}
The next point concerns the number of equilibration and measurement steps that we have to perform. It is illustrated
for an $L=16$ RE XXX chain with random bonds distributed according to Eq.(\ref{DistDis}) with $W=0.5$ in
Fig.\ref{Cz.Equil}. Here the disorder averaged mid-chain longitudinal correlation function 
\be
\label{DefCz}
C^{z}_{avg}(\frac{L}{2})={\overline{\frac{2}{L}\sum_{i=1}^{\frac{L}{2}}<S_{i}^{z}S_{i+\frac{L}{2}}^z>_{\rm{_{GS}}}}}
\ee
is plotted for different values of ($N_{eq},N_m$). The averaging is
done over $10^3$ independent samples and we observe that when the
temperature becomes low enough, even for a couple ($N_{eq},N_m$) quite
small, averaged values do not depend on the number of MC steps.  As already 
mentioned in, \cite{Sandvik.perc,SSESpins1,laflo03} we
conclude that the sample to sample variation produces larger error
bars than statistical errors, even for a number of measurement steps
$\le 100$, and in the following we will use the $\beta$-doubling
scheme with $(N_{eq},N_m)=(50,100)$ and a sufficiently large number of
samples ($\ge 10^3$).

%%%%%%%%%%%%%%%%%%%%%%%%%%%%%%%%%%%%%%%%%%%%%%%%%%%%%%%%%%%%
\begin{figure}
\bc
\epsfig{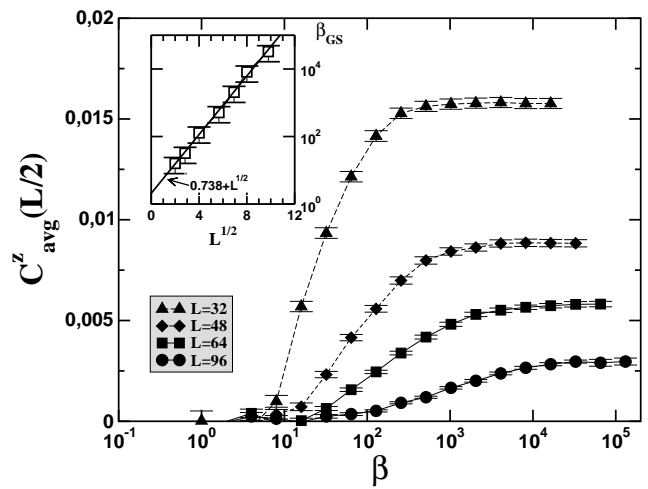} 
\caption{Test for the GS convergence of $C^{z}_{avg}(\frac{L}{2})$, 
defined by Eq.(\ref{DefCz}), vs the inverse temperature $\beta$. SSE
calculations performed on the RE XXX model for $W=0.6$ using the
$\beta$-doubling scheme with $(N_{eq},N_m)=(50,100)$. Averaging has
been done over $10^3$ different samples and the results are shown for
the 4 larger sizes $L=32, 48, 64,$ and $96$. In the inset, the GS
inverse temperature $\beta_{GS}$ (see the text for its definition) is
plotted in a log scale vs the the square root of system sizes. A
linear fit is represented by the full line.}
\label{GS.CVG}
\ec
\end{figure}
%%%%%%%%%%%%%%%%%%%%%%%%%%%%%%%%%%%%%%%%%%%%%%%%%%%%%%%%%%%%%

In order to make reliable predictions for the GS, very large $\beta$
have to be reached. This is illustrated for the RE XXX model with
disorder strength $W=0.6$ in Fig.\ref{GS.CVG}, where
$C^{z}_{avg}(\frac{L}{2})$ is plotted vs $\beta$ for different chain
sizes $L$.  We consider that the GS expectation value is obtained when
there are no statistically significant differences between the results
for $\beta_{max}=2^{n_{max}}$ and $\beta=2^{n_{max}-1}$. More precisely, our
GS convergence criterion is the following : the GS is considered reached if
the expectation value is $98 \%$ of the saturation value. Note that using such a criterion, we can
define a system size dependent temperature scale below which
the thermal expectation values are indistinguishable from GS
expectation values: $\beta_{GS}=2^{n_{max}-1}\pm 2^{n_{max}-2}$ and as
shown in the inset of Fig.\ref{GS.CVG}, we obtain for this
quantity a FS scaling of the form $\ln \beta_{GS}
\sim \sqrt{L}$ for $W=0.6$. Note that we have checked the validity of this scaling for all disorder strengths considered here. Such
a scaling is not surprising since the FS gap also obeys to a similar
law Eq.(\ref{ScalGap}).

%%%%%%%%%%%%%%%%%%%%%%%%%%%%%%%%%%%%%%%%%%%%%%%%%%%%%%%%%%%%%%%%%%%%%%%%%%%%%%%

\subsection{Spin-spin correlation functions of the random-exchange XXX model}

%%%%%%%%%%%%%%%FIG:CORRXXX%%%%%%%%%%%%%%%%%%%%%%%%%%%
\begin{figure}
\bc 
\epsfig{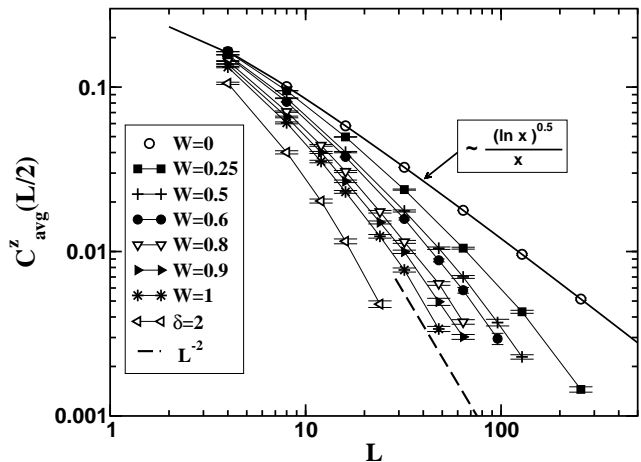} 
\caption{Averaged longitudinal correlation function $C^{z}(L/2)$ for the 
random XXX model as a function of the system size $L$ on a
log-log scale for $W=0, 0.25, 0.5, 0.6, 0.8, 0.9, 1.0$ and $\delta=2$
(top to bottom). The data, computed in the GS using SSE method and
$\beta$-doubling scheme, are averaged over more than $1000$ samples.
The data for the pure system ($W=0$, open circles) follow
$C^{z}(L/2)\propto \sqrt{\ln(L)}/L$, the dashed line with slope $-2$
is the expected asymptotic behavior according to the IRFP scenario.}
\label{fig:CORRXXX} 
\ec
\end{figure}
%%%%%%%%%%%%%%%%%%%%%%%%%%%%%%%%%%%%%%%%%%%%%%%%%%

After these careful checks of equilibration and temperature effects in
our simulations, we can analyze the SSE results obtained for the
disorder averaged longitudinal spin-spin correlation $C^{z}_{avg}$.
In order to extract the bulk value, we compute this quantity at
mid-chain and perform the averages along the chains and over random
samples, according to Eq.(\ref{DefCz}). We consider in the following
the RE XXX Hamiltonian
\be
\label{REXXX}
{\mathcal{H}}_{\rm{RE}}^{XXX}=\sum_{i=1}^{L}
\Bigl[J(i)(S_i^xS_{i+1}^x+S_i^yS_{i+1}^y+S_i^zS_{i+1}^z\Bigr],
\ee
with $J(i)$ random AF couplings taken from the $W$-dependent
distribution Eq.(\ref{DistDis}). We have also used the more singular
distribution ${\mathcal{P}}(J)=\delta J^{-1+\delta^{-1}}$ if $J\le 1$
and $0$ otherwise, with $\delta=2$.  Such a distribution is, a priori,
closer to the IRFP and therefore we expect the asymptotic behavior
$C^{z}_{avg}(\frac{L}{2})\sim L^{-2}$ to become visible already for
not too large system sizes. Indeed, this is what we can see in
Fig.\ref{fig:CORRXXX}, where $C^{z}_{avg}(\frac{L}{2})$ is plotted versus
$L$ for different disorder strengths. The crossover phenomena, already
mentioned for the RE XX case, is also clearly visible but from $16$
sites the RSP behavior is recovered for the $\delta=2$ case.  For
weaker disorder, the asymptotic behavior is visible only for larger
distances and an analysis analogous to the one we have performed for the
XX chain is necessary in order to extract a disorder-dependent
crossover length scale $\xi$.  In the pure XXX case the
exponent in Eq.(\ref{DefCor}) is $\eta_z =1$, but
logarithmic corrections have to be taken into account \cite{Affleck89}
\be
\label{Czpure}
C^{z}(r) \propto (-1)^{r}\frac{\sqrt{\ln r}}{r},
\ee
with which our numerical data for $W=0$ agree (see
Fig.\ref{fig:CORRXXX}).

%%%%%%%%%%%%%%%FIG:SCALXXX%%%%%%%%%%%%%%%%%%%%%%%%%%%
\begin{figure}
\bc 
\epsfig{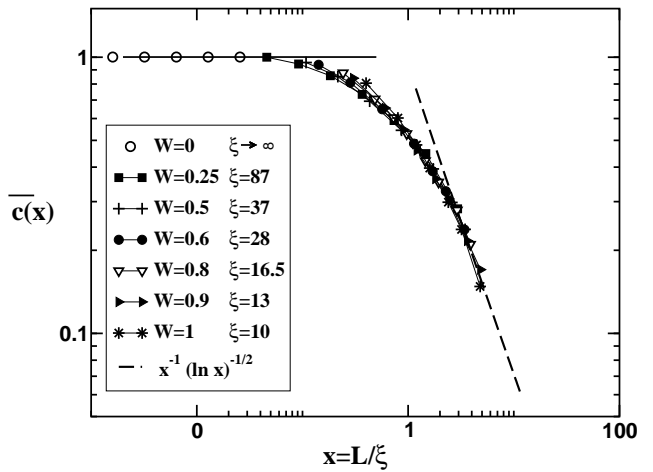} 
\caption{Scaling plot according to
  Eq. (\ref{ScalingXXX}) for the data of the Fig.\ref{fig:CORRXXX}
  with $\xi=87$, $37$, $28$, $16.5$, $13$, $10$ for $W=0.25$, $0.5$,
  $0.6$, $0.8$, $0.9$ and $1.0$, respectively. The full line stands
  for the pure behavior and the dashed line is the expected
  asymptotic behavior according to the IRFP scenario.}
\label{fig:SCALXXX} 
\ec
\end{figure}
%%%%%%%%%%%%%%%%%%%%%%%%%%%%%%%%%%%%%%%%%%%%%%%%%%

As in the XX case we expect a disorder dependent length scale $\xi$ to
govern the crossover from pure XXX behavior for $L\ll\xi$ to the
asymptotic RSP behavior visible for $L\ll\xi$. Then $C^z(L/2)$ should
obey the following scaling form:
\be
\label{ScalingXXX}
C^{z}_{avg}(\frac{L}{2})=\frac{\sqrt{\ln L}}{L}\,{\tilde{c}}(L/\xi)\;,
\ee
with ${\tilde{c}}(x)$ a scaling function that is constant in the pure
regime ($x\ll 1$) and proportional to $(x\ln^{1/2}x)^{-1}$ for $x\gg
1$ in order to reproduce the IRFP behavior
$C^{z}_{avg}(\frac{L}{2})\propto L^{-2}$ for $L\gg\xi$. In
Fig.\ref{fig:SCALXXX}, the scaling Eq.(\ref{ScalingXXX}) is shown for
the data of Fig.\ref{fig:CORRXXX}. The $W$-dependent crossover
lengths scale $\xi$ was chosen for each value for $W$ individually to
obtain the best data collapse.  $\xi(W=1)$ has been chosen such that
the crossover regime is centered around $x\simeq 1$ (i.e.\ when the
system size is of the same order of magnitude as the crossover length
scale $\xi$). In comparison with the XX results (see section 3), the
asymptotic behavior $C^{z}_{avg}\sim L^{-2}$ sets in already for
smaller system sizes. This observation is compatible with the fact
that that the disorder dependent length scale defined in Eq.(\ref{xi})
diverges much slower at the XXX point ($\xi_{XXX}\propto
{\mathcal{D}}^{-1/2}$), than at the XX point ($\xi_{XX}\propto
{\mathcal{D}}^{-1}$).

\begin{figure}
\bc 
\epsfig{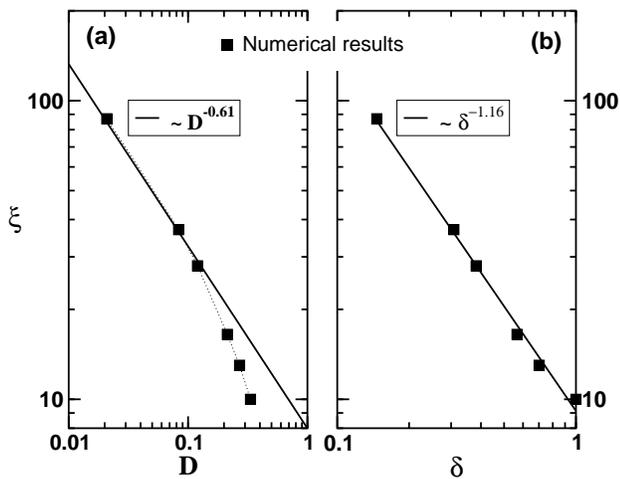} 
\caption{Disorder dependence of the crossover length scale $\xi$ of the 
random XXX chain. The full squares are the numerical estimates from
the data collapse in Fig.\ref{fig:SCALXXX}. (a) In function of the
disorder parameter ${\mathcal{D}}$, the power law fit $\xi \sim
\delta_{W}^{-0.61\pm 0.1}$ works only for weak disorder whereas in
(b), the fit $\xi \sim \delta_{W}^{-1.16\pm 0.2}$ works for the entire
range of disorder strength studied here.}
\label{fig:XiXXX} 
\ec
\end{figure}

The disorder dependence of the crossover length scale of the RE XXX
model is shown in Fig.\ref{fig:XiXXX}. For $D\to0$ it diverges with a
power law and for small disorder strengths, we can fit the data well by
$\xi({\mathcal{D}})\sim {\mathcal{D}} ^{-0.6\pm 0.1}$ [see
fig.\ref{fig:XiXXX}(a)]. As a function of $\delta$ the fit
$\xi(\delta)\sim \delta^{-1.2\pm0.2}$ is working in the whole range
of disorder strengths [see fig.\ref{fig:XiXXX}(b)]. 

The agreement of our numerical estimate of the exponent governing the
divergence of the crossover lengths with the bosonization prediction
for the localization length ($0.6\pm0.1$ versus $0.5$) is not as good
as in the XX case but still acceptable within the error bars. These
minor deviations might be due to small logarithmic corrections to
formula (\ref{xi}). This is expected since the bosonization approach
gives predictions for the RPE model, whereas the RE case considered here
is only qualitatively similar because the randomness added to the
Ising term is marginally irrelevant.\cite{Orignac}

%%%%%%%%%%%%%%%FIG:XiXXX%%%%%%%%%%%%%%%%%%%%%%%%%%%
%%%%%%%%%%%%%%%%%%%%%%%%%%%%%%%%%%%%%%%%%%%%%%%%%%

%%%%%%%%%%%%%%%%%%%%%%%%%%%%%%%%%%%%%%%%%%%%%%%%%%%%%%%%%%%%%%%%%%%%%%%%%%%%%%%
\subsection{String correlation function}

The string correlation, Eq.~(\ref{string}), was introduced to measure
hidden order in in integer spin chains where the ordinary spin-spin
correlations vanish exponentially. In the RS phase the decay of the
string correlation is expected to be described by a power law [see
Eq.(\ref{string_exp})], with a decay exponent of $\eta\sim0.382$. It
has been shown before\cite{SSESpins1,Henelius98} that the string
correlation converges particularly quickly to the expected behavior. 

In this section we begin by demonstrating yet another crossover effect 
in the random singlet phase: The RSRG calculation predicts that all 
components of the spin and string order correlations should decay with 
the same exponents although the underlying XXZ Hamiltonian is not 
rotationally invariant. This follows from the fact that the ground state 
of two S-1/2 spins coupled together by an interaction of the form 
\be
\label{XXZ}
{\mathcal{H}}=J\Bigl[S_1^xS_{2}^x+S_1^yS_{2}^y+\Delta
S_1^zS_{2}^z\Bigr]
\ee
is a rotationally invariant singlet, independently of the anisotropy
$\Delta$. So if all the spins really are bound pairwise in singlets,
then the decay of different components of the correlation functions
should be identical. However, at finite disorder strength, although the
components are found to decay with the same exponents, the prefactors
are different.\cite{Henelius98} This is due to the fact that for finite
disorder strength the strong bonds in the system are not necessarily
surrounded by much weaker bonds, which leads to fluctuations in the
singlet couplings. As the disorder strength is increased these
fluctuations should diminish and true rotational invariance should be
observed. Since the string order converges fairly quickly to the
expected random singlet exponents it is a suitable quantity to use 
to illustrate this crossover behavior.
%%%%%%%%%%%%%%%FIG:STRING0%%%%%%%%%%%%%%%%%%%%%%%%%%%
\begin{figure}
\bc 
\epsfig{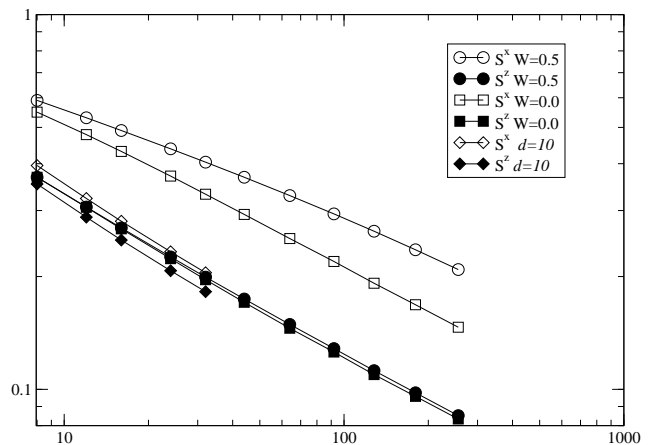} 
\caption{Exact diagonalization results for the x and z components of 
the string order at disorder strengths $W=0.5$, $W=1$ and $\delta=10$.}
\label{fig:STRING0} 
\ec
\end{figure}
%%%%%%%%%%%%%%%%%%%%%%%%%%%%%%%%%%%%%%%%%%%%%%%%%%
In Fig.~(\ref{fig:STRING0}) the $x$ and $z$ components of the string order
are shown for the XX chain with disorder parameters $W=0.5$, $W=1$ and
$\delta=10$. For flat disorder of strength $W=0.5$ the string
correlation functions  already decay with the expected RS exponent, but
the two components are quite far from each other. Increasing the
disorder strength to $W=1$ the two components approach each other,
and for a power law distribution with $\delta=10$ they are within
about 10\% of each other. Increasing $\delta$ further brings them
closer still, but it is necessary to use very high numerical precision
to get reliable data.  

\begin{figure}
\bc 
\epsfig{file=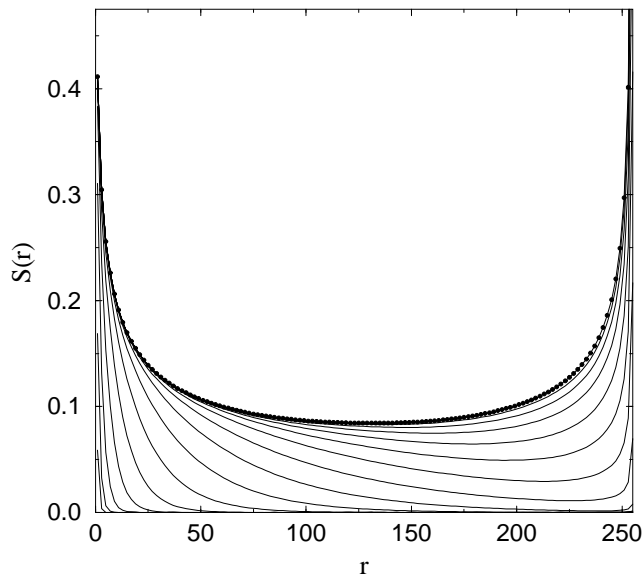,width=8.4cm,clip} 
\caption{SSE results for the string correlations of an $L=256$ XX system at 
$W=0.5$ calculated at inverse temperatures $\beta=2^n$ with $n=0,\ldots,15$. The
$\beta=2^{15}$ results are shown with solid circles; the string correlations
decrease with increasing temperature (decreasing $n$).
.}
\label{fig:qmcstring1} 
\ec
\end{figure}

In order to check the decay of the string order away from the XX point
we again use the SSE method. Here we will use chains of length $L=256$
and go to sufficiently low temperatures to observe $T\to 0$ converged
string correlations. In Fig.~\ref{fig:qmcstring1} the temperature effects are 
illustrated for an XX system at disorder strength $W=0.5$. In this case, it is
possible to obtain $T\to 0$ converged results for all distances. For
$W$ close to $1$, this would require prohibitively low temperatures,
but it is still possible to obtain well converged results up to distances
sufficiently long for observing the asymptotic RS behavior. 
Fig.~\ref{fig:qmcstring1} also illustrates that the string correlations, 
unlike spin-spin correlations, are not symmetric with respect to $r=L/2$ in 
these periodic systems. From the definition, Eq.~(\ref{string}), it is clear 
that $S(r)$ cannot be symmetric unless the total magnetization $\sum_i S^z_i = 0$.
This is the case in the ground state, where indeed the symmetry is observed.

%%%%%%%%%%%%%%%FIG:qmcstring1%%%%%%%%%%%%%%%%%%%%%%%%%%%
%%%%%%%%%%%%%%%%%%%%%%%%%%%%%%%%%%%%%%%%%%%%%%%%%%

\begin{figure}
\bc 
\epsfig{file=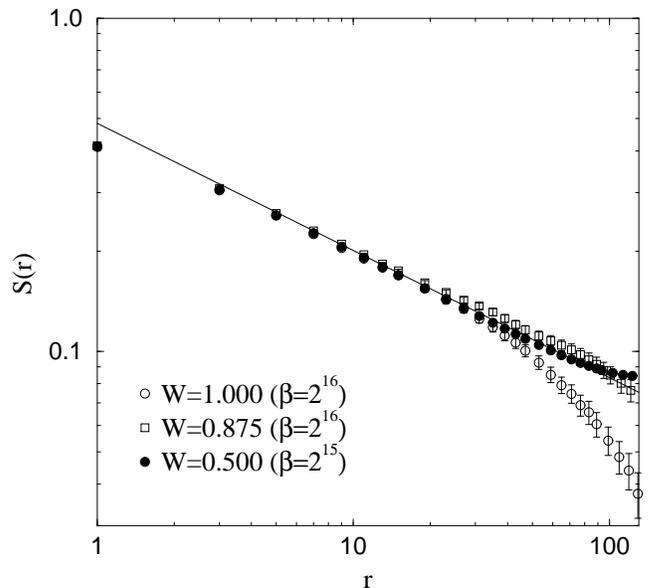,width=8.4cm,clip} 
\caption{SSE results for the string correlations of an $L=256$ XX system at 
different disorder strengths $W$, calculated at the inverse temperatures
indicated in the figure. The straight line shows the $T=0$ RS power law.}
\label{fig:qmcstring2} 
\ec
\end{figure}

\begin{figure}[h!]
\bc 
\epsfig{file=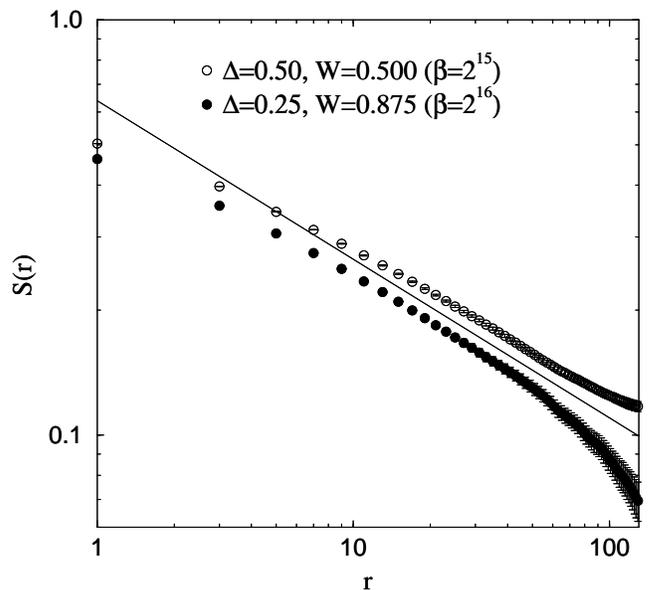,width=8.4cm,clip} 
\caption{SSE results for the string correlations of $L=256$ XXZ system at 
two combinations of Ising anisotropy $\Delta$ and disorder strength $W$.
The line shows the $T=0$ RS behavior.}
\label{fig:qmcstring3} 
\ec
\end{figure}

In Fig.~\ref{fig:qmcstring2} low-temperature results are shown at different $W$. 
Here deviations from the RS behavior due to temperature effects can be
seen for $r \agt 20$ when $W=1$, whereas deviations due to effects of the periodic
boundaries (flattening out close to $r=L/2$) can be seen at
$W=0.5$. In Fig.~\ref{fig:qmcstring3} we show similar results for the XXZ chain
for two different combinations of the Ising anisotropy $\Delta$ and 
the disorder strength $W$. In both cases RS behavior can be observed over 
a significant distance range, before temperature or boundary effects become visible
for $r \agt 50$. The very good agreement with the RS exponent provides
further evidence that the system indeed is in the RS phase for any anisotropy 
and disorder strength.

%%%%%%%%%%%%%%%FIG:qmcstring2%%%%%%%%%%%%%%%%%%%%%%%%%%%
%%%%%%%%%%%%%%%%%%%%%%%%%%%%%%%%%%%%%%%%%%%%%%%%%%
%%%%%%%%%%%%%%%FIG:qmcstring3%%%%%%%%%%%%%%%%%%%%%%%%%%%
%%%%%%%%%%%%%%%%%%%%%%%%%%%%%%%%%%%%%%%%%%%%%%%%%%

\section{Conclusion}
In this paper we have investigated numerically the spin-$\frac{1}{2}$
antiferromagnetic random-exchange XX and XXZ chains for varying
disorder strength. Using exact diagonalization calculations at the XX
point and quantum Monte Carlo SSE simulations for $\Delta \ge 0$ we
studied the ground state spin-spin and string correlation functions
for system sizes up to $L=4096$ for $\Delta=0$. With the SSE
calculations for $\Delta>0$ we went up to $L=256$ and down to very low
temperature, for instance we reached $\beta_{max}=2^{17}$ at the
random XXX point for $L=96$ and disorder strength $W=0.6$). We found clear
evidences for the asymptotically universal behavior of
the correlation functions as predicted by the RSRG analysis of Fisher
\cite{Fisher94}. The main issue of our work presented here
consists in the detailed analysis of the RG flow from the pure
instable line of XXZ fixed points toward the attractive infinite
randomness fixed point. Indeed, as we have demonstrated, such a flow is
controlled by a disorder dependent length scale $\xi$ which diverges
as the randomness approaches zero \cite{ourcomment03}. In our large scale
numerical calculations we showed that the spin-spin correlation function
is very sensitive to such crossover effects whereas the string order
converges more rapidly to its asymptotic RS value. Nevertheless the
string order also displays a crossover phenomena, visible not in 
the decay  exponents as in the spin-spin case but rather in the
prefactors.

The spin-spin correlation function as well as the stiffness display a
clear crossover from the pure behavior to the IRFP behavior as predicted
by the RSRG. The crossover length scale, extracted from numerical
data, is shown to diverge as $\xi\sim{\mathcal D}^{-\gamma}$. Our
estimates for the exponent $\gamma\approx1.0$ agrees very well
within the error bars with the localization length
exponent calculated within an analytical bosonization approach
\cite{Doty92}. However, as the bosonization approach is only valid for
a disorder that is not too strong, our estimates for the crossover
length scale $\xi({\mathcal D})$ and for the localization length
$\xi^{*}({\mathcal D})$ both deviate (in a perfectly similar way) from
the predicted behavior described by Eq.(\ref{xi}) when the disorder
strength increases. For any strength of randomness, we found a better
parameter to describe crossover as well as localization
effects. Indeed, using the variance of the logarithm of the random
couplings, $\delta$ given by Eq.(\ref{delta}), our estimates for the
crossover length scale fits in the 
whole range of disorder strengths considered here very well the form 
$\xi(\delta)\sim \xi^*(\delta) \propto \delta^{-\Phi}$ 
with $\Phi=\frac{2}{3-2 K}$. It would be
interesting to check such a $\delta$ dependence of $\xi$ or $\xi^*$
also for $\Delta \neq 0$ or $1$. The connection between crossover and
localization effects has been clearly demonstrated here and has
motivated further studies of the localization in 1D \cite{Stif03}.

Whereas the models we have studied are described by the IRFP for any
strength of the disorder, several disordered magnetic systems require
a critical value of randomness to display universal RSP features. For
instance, gaped systems like the spin-$1$ chains or spin-$\frac{1}{2}$
$n$-legs ladders are not unstable with respect to the introduction of
weak disorder and a precise identification of the critical disorder
${\mathcal D}_c$ might be easier if one considers the divergence of
$\xi$ when the disorder strength approaches the critical value
${\mathcal D} \to {\mathcal D}_c$.

NL would like to acknowledge stimulating discussions with E. Orignac.
The work of NL and HR was financially supported by the Deutsche
Forschungsgemeinschaft (DFG) and by the European Community's Human
Potential Programme under contract HPRN-CT-2002-00307, DYGLAGEMEM. 
A part of the simulations have been performed on parallel supercomputers at IDRIS (Orsay, France).

%%%%%%%%%%%%%%%%%%%%%%%%%%%%%%%%%%%%%%%%%%%%%%%%%%%%%%%%%%%

%\end{multicols}


\begin{thebibliography}{}

\bibitem{Bethe31} H.~Bethe, Z. Physik {\bf 38}, 441 (1931).

\bibitem{spins3.2} K. Hallberg, X. Q. G. Wang, P. Horsch, and A. Moreo, Phys. Rev. Lett {\bf{76}}, 4955 (1996).

\bibitem{Haldane83} F. D. M. Haldane, Phys. Lett. {\bf{93A}}, 464 (1983).

\bibitem{Mermin66} N. D. Mermin and H. Wagner, Phys. Rev. Lett. {\bf{17}}, 1133 (1966).

\bibitem{Igloi00} F. Igloi, R. Juh\'asz, and H. Rieger, Phys. Rev. B {\bf{61}}, 11552 (2000).

\bibitem{Doty92} C.Doty and D. S. Fisher, Phys. Rev. B {\bf{45}}, 2167 (1992). 

\bibitem{Fisher94} D. S. Fisher, Phys. Rev. B {\bf{50}}, 3799 (1994).

\bibitem{noteXXZ} When $J_{perp}(i)$ and $J_{z}(i)$ are independent random variables, Fisher has established a phase diagram in Fig.5 of \cite{Fisher94}.

\bibitem{HigherSpins} G. Refael, S. kehrein, and D. S. Fisher, Phys. Rev. B {\bf{66}}, R060402 (2002); K. Damle and D. Huse, Phys. Rev. Lett. {\bf{89}}, 277203 (2002); E. Carlon et al., preprint cond-mat/0301067.

\bibitem{Debate} K. Hida, Phys. Rev. Lett. {\bf{83}}, 3297 (1999); K. Yang and R. A. Hyman, Phys. Rev. Lett. {\bf{84}}, 2044 (2000); A. Saguia, B. Boechat, and M. A. Continetino, Phys. Rev. Lett. {\bf{89}}, 117202 (2002).

\bibitem{SSESpins1} S. Bergkvist, P. Henelius, and A. Rosengren, Phys. Rev. B {\bf{66}}, 134407 (2002).

\bibitem{MDH} S. K. Ma, C. Dasgupta, and C-K Hu, Phys. Rev. Lett. {\bf{43}}, 1434 (1979); C. Dasgupta and S-K MA, Phys. Rev. B {\bf{22}}, 1305 (1980).

\bibitem{Haasetal} S.~Haas, J.~Riera, and E.~Dagotto, Phys. Rev. B {\bf{48}}, R13174 (1993).

\bibitem{Henelius98} P.~Henelius and S.~M.~Girvin, Phys. Rev. B {\bf{57}}, 11457 (1998).

\bibitem{noteXX} Note that for the related model of the random transverse-field Ising model, FFED calculations have also confirmed the universal behavior on FS clusters ($L_{max}=128$ spins) : see Ref \cite{Young96}.

\bibitem{Young96} A. P. Young and H. Rieger, Phys. Rev. B {\bf{53}}, 8486 (1996).

\bibitem{Sigrist99}  T. Hikihara, A. Furusaki, and M. Sigrist, Phys. Rev. B {\bf{60}}, 12116 (1999).

\bibitem{Todo99}
S.~Todo, K.~Kato, and H.~Takayama, in {\it{Computer Simulation Studies
in Condensed-Matter Physics XI}} (Springer-Verlag Berlin Heidelberg,
1999), pp. 57-61.

\bibitem{Stolze2002}
K. Hamacher, J. Stolze, and W. Wenzel, Phys. Rev. Lett. {\bf 89}, 127202 (2002).

\bibitem{ourcomment03} N. Laflorencie and H. Rieger, Phys. Rev. Lett {\bf 91}, 229701 (2003).

\bibitem{Stolze96}
H.~R\"{o}der, J.~Stolze, R.~Silver, and G.~M\"{u}ller, J. Appl. Phys {\bf{79}}, 4632 (1996).

\bibitem{Giamarchi88} T. Giamarchi and H. J. Schulz, Phys. Rev. B {\bf{37}}, 325 (1988).  

\bibitem{Fisher92.95} D. S. Fisher, Phys. Rev. Lett. {\bf{69}}, 534 (1992); D. S. Fisher, Phys. Rev. B {\bf{51}}, 6411 (1995).

\bibitem{review00} For a review on Luttinger liquid, see H. J. Shulz, G. Cuniberti, and P. Pieri, {\it{Fermi liquids and Luttinger liquids}}, in {\it{Field Theories for Low-Dimensional Condensed matter Systems}} (Springer, 2000).   


\bibitem{Orignac} E. Orignac, Private communication.

\bibitem{anderson} E. Abrahams, P. W. Anderson, D. C. Licciardello and  T. V. Ramakrishnan, Phys. Rev. Lett. {\bf {42}}, 673 (1979).

\bibitem{Poilblanc94} G. Bouzerar, D. Poilblanc, and G. Montambaux, Phys. Rev. B {\bf{49}}, 8258 (1994).

\bibitem{Runge94} K. J. Runge and G. T. Zimanyi, Phys. Rev. B {\bf{49}}, 15212 (1994).

\bibitem{Urba03} L. Urba and A. Rosengren, Phys. Rev. B {\bf{67}}, 104406 (2003); in this study, DMRG calculations have been performed on system up to $50$ sites.

\bibitem{Stif03} N. Laflorencie and H. Rieger, in preparation.

\bibitem{Lieb61} E. Lieb, T. Schulz, and D. Mattis, Ann. Phys. (NY) {\bf{16}}, 407 (1961).

\bibitem{HuseTransport} For a study of dynamics and transport properties at non-zero frequency in random spin chains : see O. Montrunich, K. Damle, and D. Huse, Phys. Rev. B {\bf{63}}, 134424 (2001).

\bibitem{TBC} The twisted boundary conditions are given by $S^z_{L+1}=S^z_1,\ \  S^{\pm}_{L+1} = S^{\pm}_1 e^{\pm i\varphi}$.

\bibitem{Wallin94} M. Wallin, E.~S.~S\o rensen, S.~M.~Girvin, and A.~P.~Young,
Phys. Rev. B {\bf 49}, 12115 (1994). 

\bibitem{noteSti} This value $\frac{1}{\pi}$ is strictly valid in the thermodynamic limit, but the FS corrections are quite small $\propto L^{-2}$ : see N. Laflorencie, S. Capponi, and E. S. S\o rensen, Eur. Phys. J. B {\bf{24}}, 77 (2001).       

\bibitem{noteNum} Actually for a sample of size $L$, we have to solve an eigenvalues problem twice on a  $L \times L$ matrix to obtain the spin stiffness (once at $\varphi=0$ and once at $\varphi=\delta_\varphi$) whereas the evaluation of the mid-chain spin-spin correlation function requires to solve eigenvalues and eigenvectors problem on a $L \times L$ matrix and then requires to calculate a $\frac{L}{2} \times \frac{L}{2}$ determinant.

\bibitem{Sandvik.ref}
A.~W~Sandvik, Phys. Rev. B {\bf{59}}, R14157 (1999).

\bibitem{SandvikDirect02} O. F. Syljuasen and A. W. Sandvik, Phys. Rev. E {\bf{66}}, 046701 (2002).

\bibitem{Syljuasen02} O. F. Syljuasen, Phys. Rev. E {\bf{67}}, 046701 (2003).

\bibitem{Alet03} F. Alet, S. Wessel, and M. Troyer, 
preprint cond-mat/0308495. 

\bibitem{Sandvik.perc}
A.~W~Sandvik, Phys. Rev. B {\bf{66}}, 024418 (2002).

\bibitem{laflo03} N. Laflorencie, D. Poilblanc, and A. W. Sandvik, preprint cond-mat/0308334.

\bibitem{NoteTodo} S. Todo, private communication.

\bibitem{Affleck89} I. Affleck, D. Gepner, H. J. Schulz, and T. Ziman  J. Phys. A {\bf{22}}, 511 (1989).
\end{thebibliography}
\end{document}